\documentclass[12pt,a4paper]{article}

\usepackage[nosort]{cite}
\usepackage{a4wide}
\usepackage{amssymb}
\usepackage{epsfig}
\usepackage{aip}
\usepackage{amsbsy}
\usepackage{color}

\begin{document}

\title{{\bf
Phonon-driven ultrafast exciton dissociation at donor-acceptor polymer heterojunctions
}}

\pagestyle{plain}
\vskip 0.1 cm
\author{ \\
       {\normalsize \hspace*{-0.5cm}Hiroyuki Tamura,$^{\dag}$
       John G. S. Ramon,$^{\ddag}$
       Eric R. Bittner,$^{\ddag}$ and
       Irene Burghardt$^{\dag}$}
\vspace {0.1cm}
\\{\it\small $^{\dag}$D\'epartement de Chimie, 
Ecole Normale Sup\'erieure,}
\\{\it\small 24 rue Lhomond, F--75231 Paris cedex 05, France}
\\
\vspace {-0.4cm}
\\{\it\small 
$^{\ddag}$Department of Chemistry and Texas Center for Superconductivity,}
\\{\it\small University of Houston, Houston, Texas 77204, USA}
\vspace {0.2cm}
}

\maketitle

\begin{abstract}
A quantum-dynamical analysis of phonon-driven exciton dissociation
at polymer heterojunctions is presented, using a hierarchical
electron-phonon model parameterized for three electronic states and 24
vibrational modes. Two interfering decay pathways are identified:
a direct charge separation, and an indirect pathway via an intermediate
bridge state. Both pathways depend critically on the dynamical interplay
of high-frequency C=C stretch modes and low-frequency ring-torsional
modes. {The ultrafast, highly non-equilibrium
dynamics is consistent with time-resolved spectroscopic observations}.

\vspace*{1.0cm}

PACS numbers: 78.55.Kz,78.66.Qn,31.70.Hq,31.50.Gh

\end{abstract}

\sloppy

\baselineskip=23pt

\vspace*{0.2cm}

Semiconducting $\pi$-conjugated polymers are promising low-cost and
flexible materials for electronic devices such as
organic light-emitting diodes (OLEDs)
and solar cells [\citen{natu,PRB99,SR06}].
Many of the unique properties of these materials
stem from the combination of the molecular structure of their
building blocks and the extended nature of the $\pi$-system. Thus,
excitonic states, i.e., electron-hole quasi-particle states
that extend over the polymer strands, play a central
role for the optoelectronic properties [\citen{SR06}].
Exciton formation can result from polaron recombination, e.g., in
OLEDs, or else from the primary photoexcitation step, e.g., in
photovoltaic systems.

\vspace*{0.2cm}

A key feature of organic semiconductors
is the strong coupling between the elementary electronic processes
and the material's phonon modes [\citen{SR06,Tretiak02}] -- much stronger 
than in inorganic semiconductors. This is underscored by experimental
observations of sub-picosecond scale
coherent nuclear motions following
photoexcitation [\citen{LCBS03}]. 
Further, the ultrafast nature of the electronic
decay processes [\citen{Ketal93,Metal04,Aietal06}] 
suggests the presence of coherent vibronic coupling mechanisms,
possibly  determined by conical intersection topologies -- very
similar to the photophysics of polyatomic
molecules and to Jahn-Teller effects in 
solids [\citen{LVC1,CI}]. These observations
call for a microscopic-level,
quantum-dynamical interpretation of the basic charge and
energy transfer processes.

\vspace*{0.2cm}

In this Letter, we focus on the role of electron-phonon coupling
in the charge transfer processes at bulk polymer
heterojunctions [\citen{heterojunction}]; these 
provide conditions for an extremely efficient charge separation
at the interface between different phase-segregated polymers. 
The primary excitation is a photogenerated exciton stabilized
by the electron-hole Coulombic interaction (with a typical binding energy of
$\epsilon_B \sim$ 0.5 eV [\citen{PRB99}]).
{Due to the highly folded interfacial area in bulk heterojunctions,
the exciton has a high
probability of reaching the interface within the diffusion
length (typically $\sim$ 20 nm). The exciton decay towards a
charge-separated state (``exciplex'') is largely determined 
by molecular-level electronic interactions at the interface.
To a first approximation, }the efficiency of
charge generation depends on
the ratio between $\epsilon_B$ and the band offset between
the two polymer species [\citen{heterojunction,BR06}].
Recent time-resolved
photoluminescence studies have shown that the exciton decay
-- and possible exciton regeneration --
fall into a (sub)picosecond
regime [\citen{Metal04,Aietal06}].
While the elementary processes at the polymer interface are
crucial for the device function and optimization, these processes are
not well characterized as yet on the microscopic side.
Against this background, the purpose
of this Letter is to present a realistic quantum-dynamical model
involving several electronic states and an explicit representation
of the phonon distribution.

\vspace*{0.2cm}

We specifically consider a heterojunction composed of the 
poly[9,9-dioctylfluorene-co-bis-N,N-(4-butyl-phenyl)-bis-N,N-phenyl-1,4-phenylene\-di\-amine] (TFB) 
and poly[9,9-dioctylfluorene-co-benzothiadiazole] (F8BT)
polymer components [\citen{Metal04,BR06,RB06,RB07}].
At the TFB:F8BT heterojunction, 
$\epsilon_B$ is found to be similar to the band offset,
so that both exciton decay and regeneration phenomena are expected
to occur. 
Semi-empirical calculations were used to identify the relevant
electronic states and
parameterize the vibronic coupling Hamiltonian used below (see Eq.\ (1))
for 24 explicit phonon modes
covering the high-frequency (C=C stretch) and low-frequency
(ring-torsional) phonon branches [\citen{RB06,PB06,TB07}].
The exciton state (XT), which is the lowest excited state 
with significant oscillator strength, mainly has electron-hole
density on the F8BT moiety, while the lowest-lying 
interfacial charge transfer state (CT)
corresponds to a TFB$^+$-F8BT$^-$ charge separated state.
{(However, the CT state is not purely polaronic since
it does carry oscillator strength to the ground state through mixing with
excitonic configurations on the F8BT moiety).}
Among the manifold of remaining states, one is found to exhibit a
strong coupling to the XT state [\citen{RB06}];
this state will be considered as an intermediate state (IS)
in the following, see Fig.\ 1.

\vspace*{0.2cm}

To interpret the nonadiabatic decay of the exciton state, we apply 
a three-state vibronic coupling
model in conjunction with highly accurate 
wavepacket propagation techniques, i.e., the multiconfiguration
time-dependent Hartree (MCTDH) method [\citen{MCTDH}].
In particular, we use a
linear vibronic coupling (LVC) model [\citen{LVC1}]
which is appropriate in view of
the moderate nuclear displacements during the process. The
$N$-mode LVC
Hamiltonian (here, $N$=24) reads as follows,
using mass and frequency weighted coordinates and atomic units ($\hbar$ = 1),
\begin{eqnarray}
  \label{HLVC}
  \boldsymbol{H} =
  \sum_{i=1}^N 
  \frac{\omega_i}{2}
  \biggl( p_i^2\, +\, x_i^2 \biggr) {\bf 1} + 
  \sum_{i=1}^N 
  \left( \begin{array}{ccc}
    \kappa_i^{(1)} x_i &  \lambda_i^{(12)} x_i &  \lambda_i^{(13)} x_i 
    \\  \lambda_i^{(12)} x_i  &  \kappa_i^{(2)} x_i &  \lambda_i^{(23)} x_i
    \\  \lambda_i^{(13)} x_i  &  \lambda_i^{(23)} x_i  &  \kappa_i^{(3)} x_i   
  \end{array} \right)
  + \boldsymbol{C}
\end{eqnarray}
Here, $\boldsymbol{1}$ and $\boldsymbol{C}$ denote 
the unit matrix and a coordinate independent constant matrix, respectively;
the $\omega_i$, $p_i = -i{\partial}/{\partial x_i}$, and $x_i$ 
are the frequencies, momenta, and displacements along the
vibrational normal modes. The diagonal and off-diagonal
potential terms correspond to the diabatic potentials and couplings.
The model Eq.\ (1) allows for the presence of conical intersections
{between pairs of electronic states
at nuclear configurations where the diabatic coupling vanishes and the
adiabatic states become degenerate [\citen{LVC1}].}

\vspace*{0.2cm}

We further apply a recently developed
effective-mode representation of the LVC
model [\citen{CGB05}], which separates the
Hamiltonian into effective-mode vs. residual-mode parts,
\begin{eqnarray}
\boldsymbol{H} = \boldsymbol{H}_{\rm eff} + \boldsymbol{H}_{\rm res}
\end{eqnarray}
By an orthogonal coordinate transformation, $\boldsymbol{X}
= \boldsymbol{T} \boldsymbol{x}$,
a subset of effective modes is generated which subsume all
vibronic coupling terms of an electronic $N_s$-state system
into $\frac{1}{2} N_s(N_s+1)$ effective modes [\citen{CGB05}].
This construction, which was previously applied to two-state
systems (three effective modes) [\citen{TB07,CGB05,GKC07}],
is here extended for the first time to three electronic states
(six effective modes),
\begin{eqnarray}
  \label{Heff}
  \boldsymbol{H}_{\rm eff} & = &
  \sum_{i=1}^{6} \frac{\Omega_i}{2} ({P}_i^2 + {X}_i^2) \boldsymbol{1} +
  \sum_{i=1}^{6} 
  \left( \begin{array}{ccc}
    (K_i + D_i) X_i       &  \Lambda_i^{(12)} X_i  &  \Lambda_i^{(13)} X_i 
    \\  \Lambda_i^{(12)} X_i  &   (K_i - D_i) X_i  &  \Lambda_i^{(23)} X_i 
    \\  \Lambda_i^{(13)} X_i  &  \Lambda_i^{(23)} X_i  &  K_i^{(3)} X_i   
  \end{array} \right) 
  \nonumber \\
  \mbox{} &  & \hspace*{0.0cm} 
  + \sum_{i=1}^{6} \sum_{j=i+1}^{6}
  d_{ij} \biggl(
  {P}_i {P}_j + {X}_i {X}_j \biggr)
  \boldsymbol{1} +
  \boldsymbol{C} 
\end{eqnarray}
where the parameters and modes $\boldsymbol{X}$ are defined
by extension of the two-state construction described in
Refs.\ [\citen{TB07,CGB05}].
Two of the six effective modes (i.e., $X_1$ and $X_2$)
are chosen as topology-adapted modes which span the branching
plane [\citen{AXR91}] for a given pair
of electronic states (here, states 1 and 2), i.e., the
plane along which the
degeneracy at the conical intersection is lifted, see Fig. 2. 
Several of the coupling constants in Eq.\ (3), i.e., 
the $\Lambda_i^{(12)}$ ($i \ge 2$), $D_i$ ($i \ge 3$),
$K_i$ ($i = 4,5,6$), $\Lambda_i^{(12)}$ ($i = 5,6$), and
$\Lambda_6^{(13)}$, are zero by construction. 

\vspace*{0.2cm}

The residual modes, $X_i$, $i = 7, \ldots N$, do not couple
directly to the electronic
subsystem, but couple bilinearly to the effective modes.
By analogy with our recent two-state analysis [\citen{TB07}],
$\boldsymbol{H}_{\rm res}$ can be transformed to a band-diagonal
structure, yielding the $n$th-order
Hamiltonian
\begin{eqnarray}
  \label{Hn}
  \boldsymbol{H}^{(n)} = \boldsymbol{H}_{\rm eff} +  
  \sum_{l=1}^{n}\, {\boldsymbol{H}_{{\rm res}}^{(l)}}
\end{eqnarray}
where each $l$th-order residual term comprises 6 modes,
\begin{eqnarray}
     \label{Hres_n}
     \boldsymbol{H}_{{\rm res}}^{(l)} = \sum_{i=6l+1}^{6l+6}
     \frac{\Omega_i}{2} ({P}_i^2 + {X}_i^2) \boldsymbol{1} 
     + \sum_{i=6l+1}^{6l+6} \sum_{j=i-6}^{i-1}
     d_{ij} \biggl(
     {P}_i {P}_j + {X}_i {X}_j \biggr)
     \boldsymbol{1}
\end{eqnarray}
For $6 + 6n = N$, the $n$th-order Hamiltonian $\boldsymbol{H}^{(n)}$
is equivalent to the original Hamiltonian Eq. (1).

\vspace*{0.2cm}

We refer to Eqs.\ (2)-(3) in conjunction with
Eqs.\ (4)-(5) as a hierarchical electron-phonon (HEP)
model [\citen{TB07,GKC07}].
Successive orders of the HEP Hamiltonian can be shown to conserve the
moments (cumulants) of the original Hamiltonian
up to the (2$n$+3)rd order [\citen{TB07}].
Truncation of the HEP Hamiltonian 
creates a series of reduced dimensionality
models, associated with approximate $n$th-order propagators which 
accurately reproduce the dynamics of the overall system up
to a certain time. While the
lowest-order description -- $\boldsymbol{H}_{\rm eff}$ only --
can be appropriate, e.g., for a rapid passage 
through a conical intersection
region [\citen{CGB05}], the present system will be shown to
necessitate a higher-order analysis.

\vspace*{0.2cm}

The effective-mode construction for the TFB:F8BT heterojunction
shows that $\boldsymbol{H}_{\rm eff}$ is
essentially determined by high-frequency C=C stretch modes
while $\boldsymbol{H}^{(1)}_{\rm res}$ is determined
by low-frequency ring torsional modes.
This is analogous to our previous analysis of a two-state
XT-CT model [\citen{TB07}]. Importantly, the low-frequency modes will be shown
to play a key role in generating the observed ultrafast exciton 
decay.

\vspace*{0.2cm}

Fig.\ 2 shows the diabatic XT, CT, and IS potential energy surfaces (PES),
for a cut through the XT-CT branching plane defined in terms
of the effective modes $(X_1, X_2)_{\rm XT,CT}$, and an alternative
representation involving the XT-IS branching plane spanned by
another set of effective modes $(X'_1, X'_2)_{\rm XT,IS}$.
As can be inferred from the figure, the 
exciton dissociation process is determined by
a landscape of multiple intersecting surfaces,
and the branching plane
topology is a key factor
in the nonadiabatic dynamics. Of particular
interest for the present analysis is the role of the intermediate 
state (IS) that lies above the exciton state but could be accessible due
to its strong diabatic coupling to the XT state, and due to the
significant extension of the wavepacket.

\vspace*{0.2cm}

Fig. 3a shows the population evolution obtained from MCTDH
simulations for the overall 3-state, 24-mode system
according to Eq.\ (1). An ultrafast
initial XT $\rightarrow$ CT decay takes place 
($\sim$ 50\% at 150 fs), followed by an oscillatory
behavior of the XT and CT populations which is found to
persist over the
3 picosecond observation interval.
These observations are in qualitative agreement
with the experimentally observed exciton decay and 
regeneration [\citen{Metal04}], even though
a steady overall increase of the CT population is eventually
observed in our simulation. The IS population experiences an initial
increase, but does not rise above an average of $\sim$ 10\%
beyond 500 fs.
Yet, one cannot exclude the possibility
that this state acts as a ``bridge'' facilitating the XT-CT
transfer.

\vspace*{0.2cm}

To investigate the possibility of an exciton decay pathway via
the IS state, XT $\rightarrow$ IS $\rightarrow$ CT,
we set the XT-CT diabatic coupling
artificially to zero, such that only the
indirect pathway is permitted. As can be inferred from Fig. 3b,
a subpicosecond exciton decay and concomitant rise of the CT
state population is also observed in this case, even though
the process is somewhat slower than in the fully coupled
3-state dynamics. Again, the IS population tends towards an
average of $\sim$ 10\%, i.e., the IS state facilitates the
XT-CT transition but does not act as an additional acceptor state.
Overall, the dynamics of Fig 2a is therefore the result of a
superposition of the direct (XT $\rightarrow$ CT) and indirect
(XT $\rightarrow$ IS $\rightarrow$ CT) pathways. 
Given that the XT$\rightarrow$ IS transfer is endothermic (with a $\sim$ 0.2 eV 
barrier) in a
conventional kinetic picture, the efficiency of the indirect
pathway is entirely a consequence
of the quantum dynamical character of the process.

\vspace*{0.2cm}

The remaining part of the analysis addresses the HEP 
construction of Eqs.\ (2)-(5), with the twofold objective of
identifying a reduced-dimensionality picture of the dynamics, and
understanding the role of the different types of phonon modes, i.e., 
high-frequency C=C stretch modes
(constituting $\boldsymbol{H}_{\rm eff}$) vs.\ low-frequency
ring-torsional modes
(constituting $\boldsymbol{H}_{\rm res}^{(1)}$). 
As shown in Fig. 4, the high-frequency modes by themselves,
at the level of the 6-mode $\boldsymbol{H}^{(0)}$ approximation,
cannot reproduce the exciton decay (even though they
correctly reproduce the shortest-time dynamics, on the
order of 50 fs). At this level of approximation,
quantum phase coherence effects are substantially
overestimated, leading to a weak, periodic 
population transfer. By contrast, inclusion of the
low-frequency modes, at the level of the 12-mode
$\boldsymbol{H}^{(1)}$ approximation,
results in a qualitatively correct description of the
dynamics. Indeed, 
the low-frequency modes induce vibrational energy redistribution
and dephasing effects that eventually lead to an irreversible exciton
decay. 
Finally, the $\boldsymbol{H}^{(2)}$, 18-mode approximation is 
essentially identical to the exact, 
24-mode result on a $\sim$ 1 ps time
scale. A similar interpretation holds for the indirect
pathway in the absence of XT-CT coupling. 
All of these results would carry over to an initial parameterization
involving a very large number $N$ of phonon modes, e.g., thousands of
modes. While dynamical calculations according to Eq. (1)
would not be feasible for such a ``macrosystem'', the HEP model gives
a correct reduced-dimensionality description of the ultrafast
dynamics.

\vspace*{0.2cm}

{In summary, we present the results of fully quantum dynamical
calculations of a model polymer heterojunction interface. 
In particular, we highlight the role that intermediate 
states may play in the overall XT$\rightarrow$CT conversion.   
While these results go significantly beyond our earlier
interpretation for a two-state model [\citen{TB07}], the key role of
the low-frequency modes is confirmed, pointing to the
generic nature of the observed dynamical pattern for a two-band
phonon distribution (see also  
Refs.\ [\citen{Tretiak02,KBB01}] which address the spectroscopic 
signature of both types of phonon modes in phenylene based polymers). 
Our analysis highlights the coherent, quantum-dynamical
character of the process, and the importance of the
multidimensional topology. Thus, 
the branching plane of Fig. 2 replaces the
conventional Marcus parabola picture [\citen{Marcus}], and 
the HEP model provides the correct reduced-dimensionality
dynamics on short time scales.
Extensions of the present analysis concern (i) the inclusion
of temperature effects, and (ii) the improvement of the present
parameterization, along with
the generalization to multiple electronic states that potentially
appear in the role of bridge states. {The positions of
such states relative to the XT and CT states and their electronic
properties are indeed quite sensitive to the local morphology and
packing geometries of the respective polymers at the interface as
indicated by recent time-dependent density functional theory
[\citen{RB07}] and
semi-empirical investigations [\citen{Setal06}]. Consequently,
the extent to which such
intermediate states will play a role will depend on
the local morphology of the
interface and the actual physical system will be an ensemble of
individual cases.
Using this local, molecular-level information,
the HEP approach should contribute
to a realistic modeling of the highly nonequilibrium exciton
decay processes in extended $\pi$-conjugated systems. 

\vspace*{0.2cm}

We thank Andrey Pereverzev, Etienne Gindensperger, and 
Lorenz Cederbaum for constructive discussions. 
This work was supported by the ANR-05-NANO-051-02 and
ANR-NT05-3-42315 projects, by NSF grant CHE-0345324, and by the
Robert Welch Foundation. 

\newpage


\newpage

\section*{\bf Figure legends}

Fig. 1. 
(Color online)
Schematic illustration of (a) the TFB:F8BT donor-acceptor
heterojunction and (b) the photoexcitation
process and potential crossings.
The blue, red, and purple lines indicate 
the XT, CT, and IS diabatic potentials, respectively.
(Dotted lines indicate the corresponding adiabatic potentials.)
The red and purple arrows indicate the direct
XT $\rightarrow$ CT vs.\ indirect XT $\rightarrow$ IS $\rightarrow$ CT
exciton dissociation pathways.

\vspace*{0.2cm}

Fig. 2.
(Color online)
Branching-plane projections of the coupled diabatic PESs:
(a) XT, CT and IS PESs projected onto the XT-CT branching plane, and
(b) XT and IS PESs projected onto the XT-IS branching plane;
here, the wavepacket width and trajectory of the wavepacket 
center (blue arrow) are also indicated.
(These branching space projections are associated with two different
sets of effective modes $\boldsymbol{X}$.)
The white, blue, red and purple circles indicate the conical intersection,
the minima on the XT, CT and IS states, respectively.
The dotted lines indicate the avoided-crossing seam lines.

\vspace*{0.2cm}

Fig. 3.
(Color online)
Time-evolving state populations for 
(a) the full 3-state 24-mode wavepacket propagation, and
(b) a complementary 3-state 24-mode calculation where the indirect
(XT $\rightarrow$ IS $\rightarrow$ CT) pathway was selected by
setting the XT-CT diabatic coupling to zero.

\vspace*{0.2cm}

Fig. 4.
(Color online)
CT populations for the truncated HEP Hamiltonian
$\boldsymbol{H}^{(n)}$ of Eq.\ (4), for $n=0$ (6 modes),
$n=1$ (12 modes), and $n=2$ (18 modes), as compared with the
exact 24-mode result.

\newpage

\vspace*{2.0cm}

\begin{center}
\includegraphics[angle=0,width=12cm]{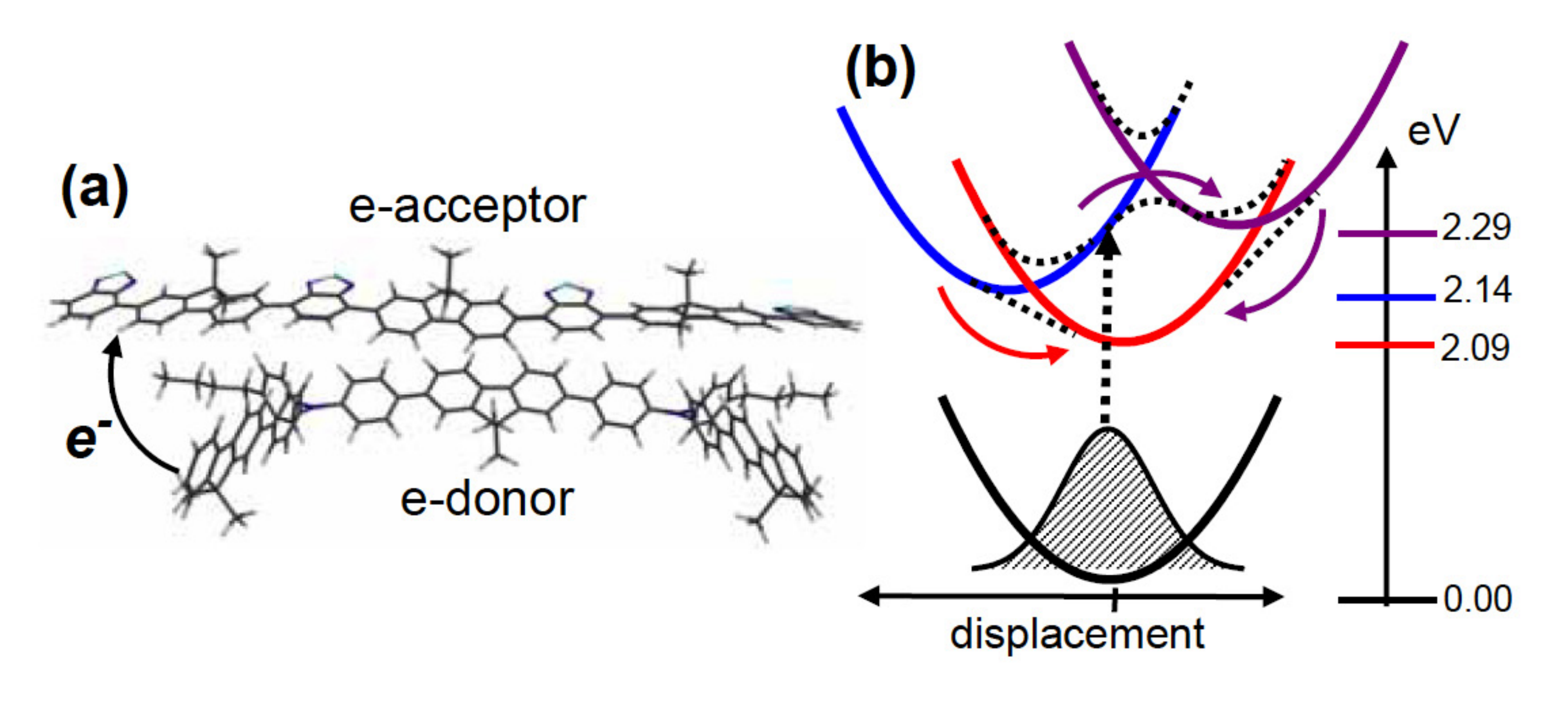}
\end{center}

\vspace*{1.0cm}

{\large\bf Fig.\ 1}

\newpage

\vspace*{2.0cm}

\begin{center}
\includegraphics[angle=0,width=14cm]{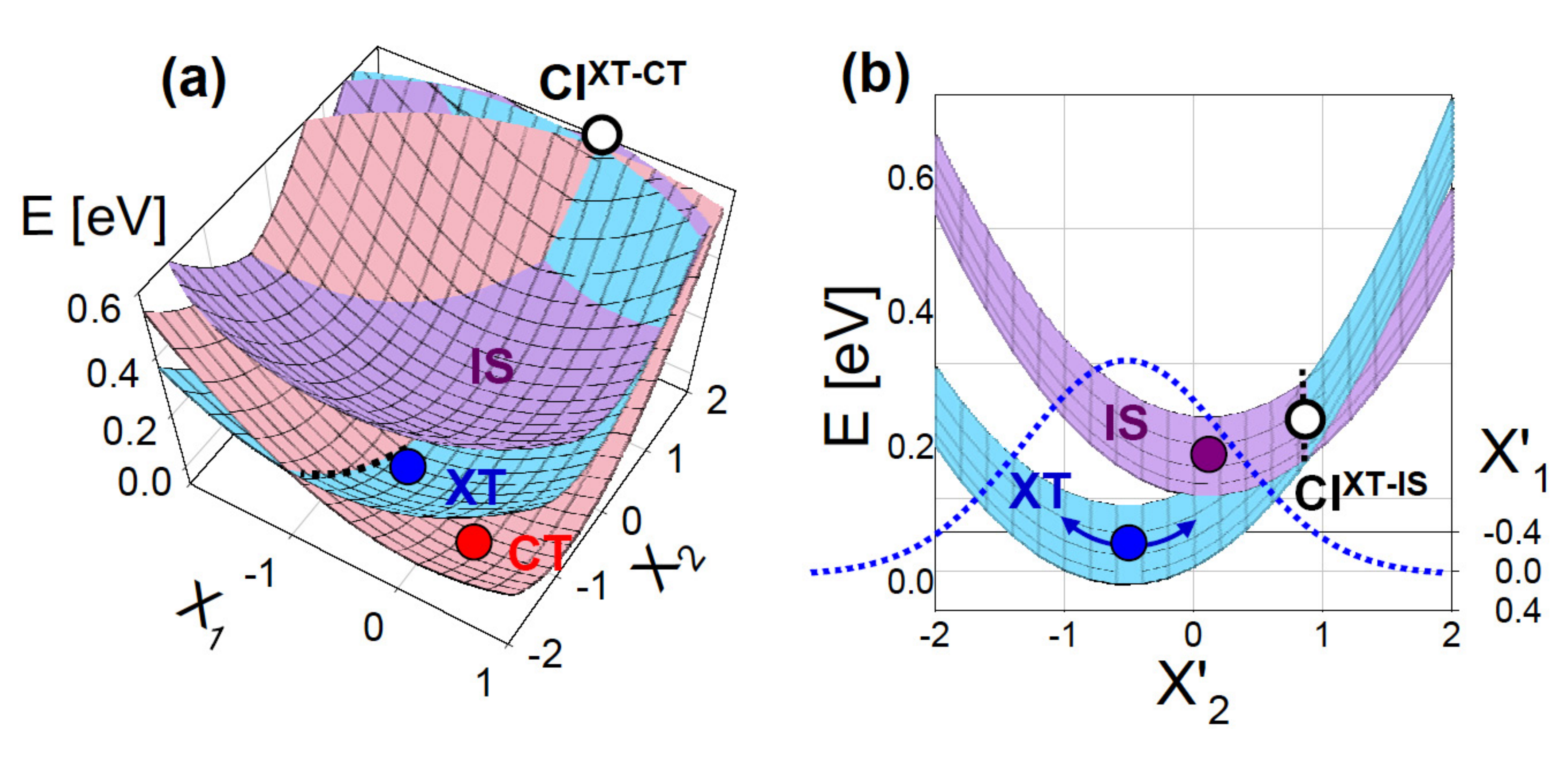}
\end{center}

\vspace*{1.0cm} 

{\large\bf Fig.\ 2}

\newpage 

\vspace*{2.0cm}

\begin{center}
\includegraphics[angle=0,width=10cm]{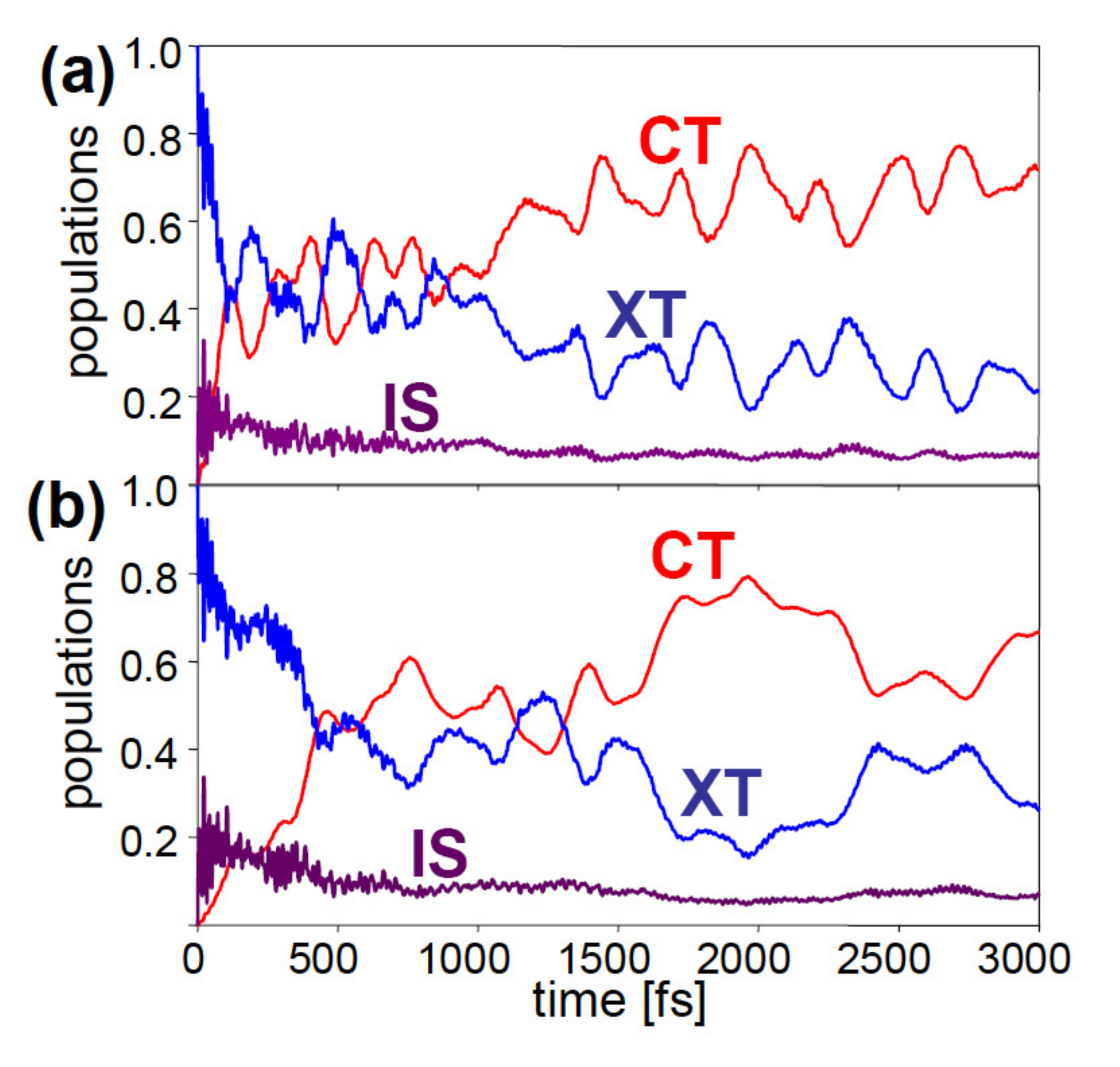}
\end{center}

\vspace*{-0.0cm} 

{\large\bf Fig.\ 3}

\newpage 

\vspace*{2.0cm}

\begin{center}
\includegraphics[angle=0,width=10cm]{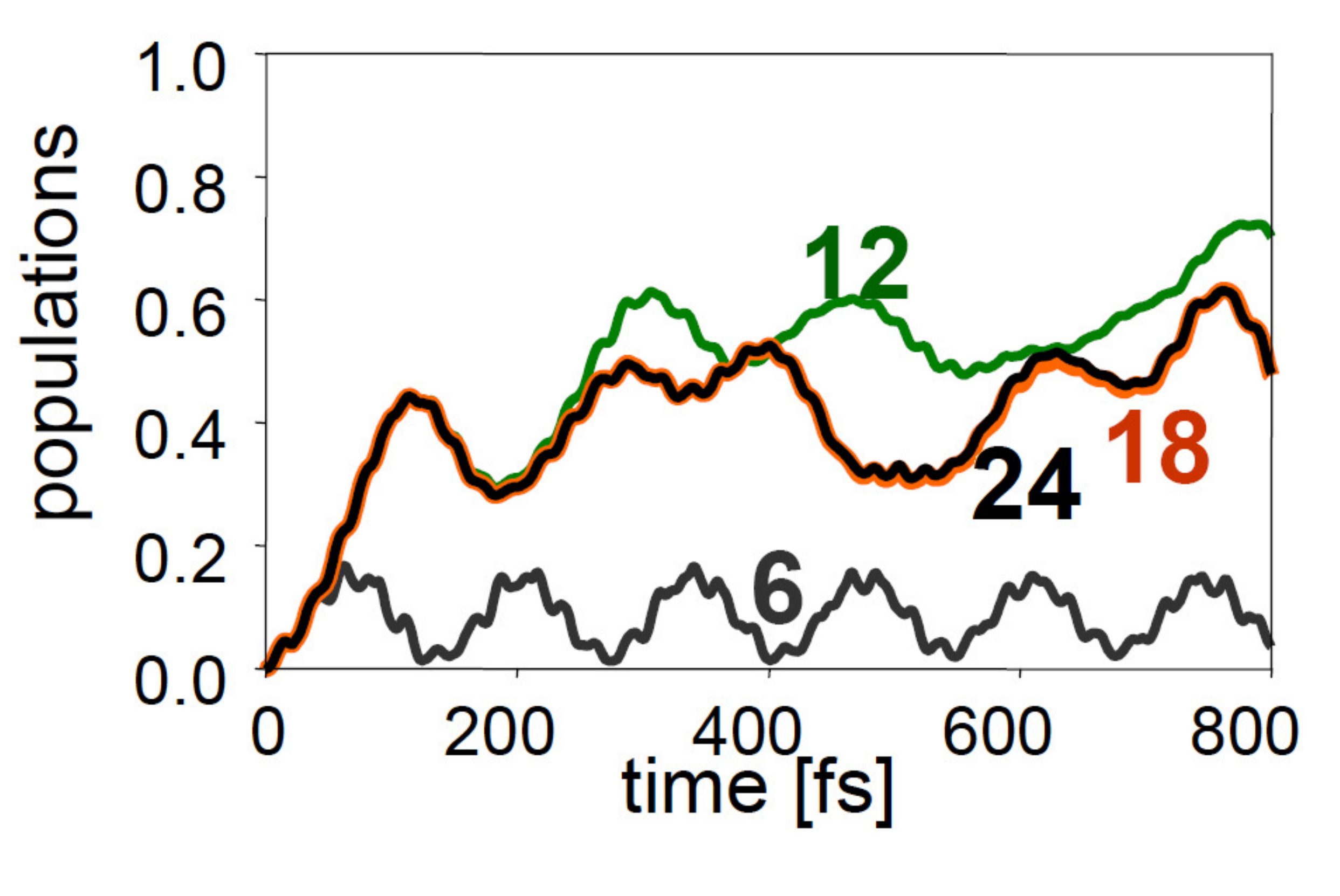}
\end{center}

\vspace*{1.0cm}

{\large\bf Fig.\ 4}

\end{document}